\documentclass[11pt]{article}
\usepackage{graphicx}
\usepackage{xcolor}
\usepackage{amsmath}

\begin{document}

\def\xslash#1{{\rlap{$#1$}/}}
\def \p {\partial}
\def \dd {\psi_{u\bar dg}}
\def \ddp {\psi_{u\bar dgg}}
\def \pq {\psi_{u\bar d\bar uu}}
\def \jpsi {J/\psi}
\def \psip {\psi^\prime}
\def \to {\rightarrow}
\def\bfsig{\mbox{\boldmath$\sigma$}}
\def\DT{\mbox{\boldmath$\Delta_T $}}
\def\xit{\mbox{\boldmath$\xi_\perp $}}
\def \jpsi {J/\psi}
\def\bfej{\mbox{\boldmath$\varepsilon$}}
\def \t {\tilde}
\def\epn {\varepsilon}
\def \up {\uparrow}
\def \dn {\downarrow}
\def \da {\dagger}
\def \pn3 {\phi_{u\bar d g}}

\def \p4n {\phi_{u\bar d gg}}

\def \bx {\bar x}
\def \by {\bar y}

%  Final revision on 16.09.2026.....

\begin{center}

\par
{\Large\bf   T-odd Effect from Two-Photon Exchange in SIDIS at Low Transverse Momentum} 
\par\vskip20pt
 J.P. Ma$^{1,2,3}$ and G.P. Zhang$^{4}$    \\
{\small {\it
$^1$ School of Physics, Henan Normal University, Xinxiang, Henan 453007,  China\\
$^2$ Institute of Theoretical Physics, P.O. Box 2735, Chinese Academy of Sciences, Beijing 100190, China\\
$^3$ School of Physics and Center for High-Energy Physics, Peking University, Beijing 100871, China\\
$^4$ Department of Physics, Yunnan University, Kunming, Yunnan 650091, China}} \\
\end{center}

\vskip 1cm
\begin{abstract}
We study T-odd effects through two-photon exchange in semi-inclusive deeply inelastic scattering in the kinematic region where 
the produced hadron has a low transverse momentum. The absorptive part of lepton-quark scattering amplitude is calculated. 
Transverse momentum dependent factorization is performed with the absorptive part. Although the absorptive part is divergent, the physical 
results, i.e., the obtained differential cross-section, is finite.  Including the T-odd effects,  the differential cross section receives new contributions  
which have different angular dependence than that obtained with one-photon exchange. Certain asymmetries can be introduced to detect the T-odd effects. Our results here 
are not only relevant to detecting new effect beyond the one-photon approximation in the process but also relevant to searching for new physics at lepton-proton colliders.  

\end{abstract}      
\vskip 5mm

\vskip40pt

\noindent 
{\bf 1. Introduction}

Lepton-proton colliders are powerful tools to explore the internal structure of hadrons as bound states of partons which are quarks and gluon. The exploration 
can reveal nonperturbative features of the strong interaction, whose fundamental theory is QCD with quark and gluons as dynamical freedoms.
The well-known scattering processes studied at lepton-colliders for the purpose are deeply-inelastic scattering(DIS) and semi-inclusive DIS(SIDIS).  
Theoretical predictions for the two scattering progresses are made with QCD factorization theorems in the form that differential cross sections 
are convolutions of perturbative coefficient functions with various parton distributions of the initial hadron and parton fragmentation functions. 
This allows 
one to extract these parton distributions from experimental results, and to learn the internal structure of hadrons.

 The QCD factorizations of DIS and SIDIS are made in the case that the interaction between the initial lepton and initial proton is approximated  as through one-photon exchange, i.e., at tree-level of QED.  Beyond the approximation, the radiative correction of QED to the case has been studied in \cite{Akushevich:2026mje,Akushevich:2023jjc,Liu:2020rvc,Cammarota:2025jyr} and references therein. The studied corrections are important 
 for extracting parton distributions precisely. In these studies, one has included the QED corrections to the lepton-photon vertex, photon propagator 
 and radiation of a real photon from the initial- and final lepton. The contribution from two-photon exchange is not included for various reasons, which are listed in e.g., \cite{Akushevich:2026mje}.

 In this work we study some particular effects of two-photon exchange in the process of SIDIS at low transverse momentum. 
 In the case of the produced hadron with low transverse momentum. Based on Transverse-Momentum-Dependent(TMD) factorization studied in \cite{CSS,Ji:2004wu}  and reviewed in \cite{Boussarie:2023izj}, the differential cross-sections can be factorized with TMD quark distributions and TMD quark fragmentation functions. Combining experimental results of the process with theoretical predictions it enables one to explore the three-dimensional inner structure of the initial hadron. The particular effects we studied here are T-odd effects from QED. These effects are absent in the approximation of one-photon exchange, but they will appear 
 beyond the approximation through two-photon exchange, where the scattering amplitude of the initial lepton and a quark from the initial hadron 
 has a nonzero absorptive part.  It is noted that the absorptive part is Infrared(I.R.) divergent. But this divergence will not appear in physical effects considered here.   

Two purposes motivate our study in this work. One is to see any new effects which are absent in the approximation of one-photon exchange, while another one is related to the search of new physics beyond the standard model.  Recently, there have been discussions in \cite{Huang:2025ljp,Wen:2024nff,Wen:2023xxc}
about detecting new interactions beyond the standard model at lepton-proton or -ion colliders like EiC in U.S.\cite{Accardi:2012qut} and EicC in China\cite{Anderle:2021wcy,Xiao:2026tbs}. 
Among them the new 
CP-violating interactions can be the most interesting ones. In general, the effects of CP-violation are expected to be small. 
Since the initial state for a scattering at lepton-proton colliders is not a CP-eigenstate,
effects from possibly new CP-violating interactions can be mixed with T-odd effects in the standard model. Therefore, to identify the true new CP-violating interactions, one needs at first to know the standard T-odd effects.  

T-odd effects from two-photon exchange have been studied in inclusive DIS  in \cite{Metz:2006pe,Metz:2012ui,Schlegel:2012ve}, where the T-odd effects 
appear as an asymmetry of single transverse spin of the initial hadron. The effect is partly related to matrix elements of the initial hadron, which 
is defined with electromagnetic field strength tensor and quark fields.  For SIDIS at low transverse momentum, the perturbative part 
of TMD factorization is essentially determined by the scattering amplitude of the initial lepton with a quark as a parton from the initial hadron. For our purpose, we will first calculate 
the absorptive part of the scattering amplitude, where two photons are exchanged between the lepton and the quark. Then we perform the TMD factorization 
at tree level of QCD for obtaining our results. The T-odd effect we study has been studied in \cite{Schlegel:2009pw} where only a result of one asymmetry is given. We will derive the complete angular distribution of the T-odd effect. 

Our work is organized as the following:  In Sect.2 we give our results of the absorptive part of the lepton-quark scattering amplitude with two-photon exchange. In Sect.3 we perform the TMD factorization for the QED T-odd effect  at tree-level of QCD. In Sect.4 we derive the angular distributions and asymmetries of T-odd effect. Sect. 5 is our summary.   An appendix is attached to give some detailed results for the scattering amplitude with two-photon exchange.   

\par\vskip20pt

\noindent 
{\bf 2. T-odd effect from two-photon exchange  in the elastic lepton-quark scattering}

\par\vskip5pt

We consider the scattering of a lepton $\ell$ with a quark:
\begin{equation} 
   \ell (k_1) + q(p_1) \to \ell (k_2) + q (p_2), 
\end{equation}
where the momentum of each particle is given in the round brackets.  The standard Mandelstam variables for the scattering are:
\begin{equation}
    s= (k_1+p_1)^2, \quad t =(p_2-p_1)^2=(k_1-k_2)^2. 
\end{equation} 
The tree-level amplitude is given by the left diagram in Fig.1, which is 
\begin{equation} 
{\cal M}_{1\gamma} (p_1,p_2) = - \frac{ e^2 e_q}{t} \bar u(k_2) \gamma^\mu u(k_1) \bar u(p_2) \gamma_\mu u(p_1)
\end{equation} 
with $e_q$ is the electric charge fraction of the quark in unit of the positron charge $e$. We will neglect the mass of leptons and quarks. 

%%%%%%%%%%%%%%%% Inset Fig. 1 here %%%%%%%%%%%%%%%%%%%%%%%%%%%%%%
\begin{figure}[hbt]
\begin{center}
\includegraphics[width=12cm]{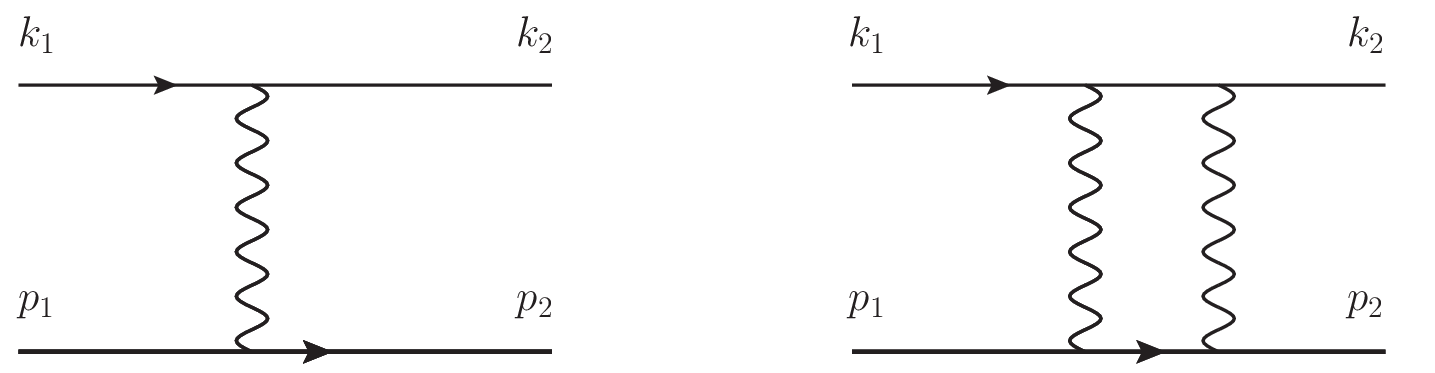}
\end{center}
\caption{Diagrams for the lepton-quark scattering with one- and two-photon exchange.  }
\label{F1}
\end{figure}

%%%%%%%%%%%%%%%% End Fig. 1 here %%%%%%%%%%%%%%%%%%%%%%%%%%%%%%
\par 
At the next-to-leading order, T-odd effect can appear from the absorptive part of the scattering amplitude which is given by the right  diagram in Fig.1, where two photons are exchanged between the lepton and the quark. The amplitude of two-photon exchange 
is given by:
\begin{eqnarray}
{\cal M}_{2\gamma} (p_1,p_2) &=& - i e^4 e_q^2 \int \frac{d^d l}{(2\pi)^d} 
    \bar u(k_2) \gamma^\beta \frac{ \gamma\cdot (k_1+l)}{(k_1+l)^2+i\varepsilon} \gamma^\alpha u(k_1) \bar u(p_2) \gamma_\beta \frac{\gamma\cdot (p_1-l)}{(p_1-l)^2+i\varepsilon} \gamma_\alpha u(p_1)
\nonumber\\
    && \frac{1}{l^2+i\varepsilon} \frac{1}{ (l+k_1-k_2)^2+i\varepsilon}. 
 \label{2PE}  
 \end{eqnarray} 
 In order to handle possible divergence we calculate the contribution in $d$-dimension with $d=4-\epsilon$. 
 
The calculation of the contribution from two-photon exchange is tedious but straightforward. We present the calculation in the Appendix. The final results can be organized into the following form with the equation of motion of a free fermion:
\begin{eqnarray}
{\cal M}_{2\gamma} (p_1,p_2)  &=& e^4 e_q^2 \biggr [ {\cal A}_1(s,t) \bar u(k_2) \gamma_\mu u(k_1) \bar u(p_2) \gamma^\mu u(p_1) 
    + {\cal A}_2(s,t) \bar u(k_2) \gamma\cdot p_1  u(k_1) \bar u(p_2) \gamma\cdot k_1 u(p_1) 
\nonumber\\
     && \quad + {\cal A}_3(s,t) \bar u(k_2) \gamma_\mu \gamma\cdot p_1 \gamma_\nu u(k_1) \bar u(p_2) \gamma^\mu \gamma\cdot k_1 \gamma^\nu  u(p_1) 
\nonumber\\
      && \quad +{\cal A}_4 (s,t) \bar u(k_2) \gamma_\mu\gamma_\alpha \gamma_\nu  u(k_1) \bar u(p_2) \gamma^\mu \gamma^\alpha \gamma^\nu u(p_1)  \biggr ], 
\label{ADEF} 
\end{eqnarray} 
where $A_i (i=1,2,3,4)$ are complex scalar functions, whose imaginary parts give the absorptive part of the amplitude. 
The imaginary parts are
\begin{align}
\text{Im}\mathcal{A}_1=& K_\epsilon\Big(\frac{\mu^2}{-t}
\Big)^{\epsilon/2}\frac{\pi}{t}\Big[
\frac{8}{\epsilon }+\frac{2t}{(s+t)^2}
\Big((2t+s)\ln\frac{-t}{s}-s-t\Big)
\Big],\notag\\
\text{Im}\mathcal{A}_2=& -K_\epsilon
\frac{8\pi}{t(s+t)^2}\Big[s+t-t\ln\frac{-t}{s}\Big],\notag\\
\text{Im}\mathcal{A}_3=& K_\epsilon
\frac{\pi}{s(s+t)^2}\Big[s+t+s\ln\frac{-t}{s}\Big],\notag\\
\text{Im}\mathcal{A}_4=&-K_\epsilon
\frac{\pi}{2(s+t)}\ln\frac{-t}{s},
\label{ImA} 
\end{align}
where 
\begin{align}
K_\epsilon=\frac{(4\pi)^{\epsilon/2}}{16\pi^2}\Gamma(2+\frac{\epsilon}{2}).
\end{align}

It is noticed that only ${\cal A}_1$ contains an I.R. singularity represented by the term with the pole  in $\epsilon= 4-d$. This divergence reflects the fact that the electromagnetic interaction is a long-range interaction mediated by massless photons.  It is also noticed that the absorptive part from  ${\cal A}_1$ is proportional to the tree-level amplitude. The physical T-odd effects are from the interference of the absorptive part with the tree-level amplitude at the considered order. Therefore,  the absorptive part from  ${\cal A}_1$ gives no contribution, or the prediction of physical T-odd effects is finite. This is similar to the T-odd effects in the case of the process $e^+e^-\to q\bar q$ studied in \cite{Kane:1978nd,Bernreuther:1992ef}.

  \par

\par\vskip20pt

\noindent 
{\bf 3.  Factorization of QED T-odd effects for SIDIS at Low Transverse Momentum}

We consider now the SIDIS process: 
\begin{equation} 
      \ell (k_1) + P(P) \to \ell (k_2) + h (P_h) + X, 
\end{equation}
where the initial proton has the momentum $P$, and the produced hadron $h$ has the momentum $P_h$. 
The standard variables for SIDIS are defined as:
\begin{equation} 
   q^\mu = k_1^\mu -k_2^\mu, \quad x_B =\frac{Q^2} {2 P\cdot q}, \quad y= \frac{P\cdot q}{P\cdot k_1}, 
  \quad   z_h =\frac{P\cdot P_h}{P\cdot q},  \quad S_{\ell P} =(k_1+P)^2, 
\end{equation}
with $Q^2=-q^2 >0$. We will neglect the mass of the initial proton because we have a very large $Q^2$ here in the process. 
The transverse momentum of $P_h$ can be defined in a frame where the spacial vector of $q$ along $-z$-direction and the initial proton moves in the $+z$-direction. The transverse momentum $P_{h\perp}$ is then perpendicular to the $z$-direction. However, $P_{h\perp}$ can be defined covariantely by the transverse metric:
\begin{equation} 
    P_{h\perp}^\mu = g_\perp^{\mu\nu} P_{h\nu}, \quad g_\perp^{\mu\nu} = g^{\mu\nu} - \frac{P^\mu \tilde q^\mu+ P^\nu \tilde q^\mu}{P\cdot q}, 
\label{TD} 
\end{equation}
with $\tilde q^\mu = q^\mu + x_B P^\mu$.  $P_{h\perp}$ is perpendicular to $q^\mu$ and $P^\mu$ simultaneously, i.e., 
$q\cdot P_{h\perp} = P\cdot P_{h\perp}=0$. We will consider the process in the kinematic region with 
$P_{h\perp} \ll Q$ and that $z_h$ is moderately large. This indicates that the hadron is produced in current fragmentation region. 

\par 
In the kinematic region we consider, the differential cross section can be factorized as a convolution of TMD quark distributions and 
TMD quark fragmentation functions with perturbative coefficient functions\cite{Ji:2004wu}.  The TMD quark distributions are defined through the density matrix:   
\begin{eqnarray}
{\mathcal M}_{ij}(x, k_{\perp})   &=&  \int \frac{ d\xi^-d^2\xi_\perp } {(2\pi)^3} e^{-i \xi \cdot  k}
\langle P(P,S) \vert   \bar q_j (\xi) 
  q_i  (0) \vert P(P,S) \rangle \biggr\vert_{\xi^+ =0} ,
 \label{DENM}
\end{eqnarray}
where $q$ is the quark field and $i$ or $j$ is its color-and Dirac index. Here for the definition we use the light-cone coordinate system, in which a
vector $a^\mu$ is expressed as $a^\mu = (a^+, a^-, \vec a_\perp) =
((a^0+a^3)/\sqrt{2}, (a^0-a^3)/\sqrt{2}, a^1, a^2)$ and $a_\perp^2
=-(a^1)^2-(a^2)^2$. In this system, we introduce two light cone vectors: $l^\mu =(1,0,0,0)$ and $n^\mu=(0,1,0,0)$. The initial proton moves along the $z$-direction with the momentum $P^\mu =(P^+, P^-, 0,0)= P^+l^\mu + P^-n^\mu$, its polarization is given by the spin vector $S^\mu$ which can be written as:
\begin{equation} 
 S^\mu = S_L \frac{ P\cdot n l^\mu -P\cdot l n^\mu}{M} + S^\mu_\perp, \ \ \ \  S_L = M\frac{n\cdot S}{n\cdot P}, 
\end{equation} 
where $M$ is the mass of the initial proton. 
For $\vec S_\perp =0$ one can show that $S_L$ is just the helicity of the hadron. 
In Eq.(\ref{DENM}) the quark as a parton carries the $+$-component $k^+$ as $xP^+$ and the transverse momentum $k_\perp^\mu$. The $-$-component is integrated out.

The density matrix ${\mathcal M}$ can be decomposed into various TMD parton distributions.
The decomposition has been studied in \cite{TMDBM,TMDGMS,BDGM}. At leading power or leading twist, the decomposition is: 
\begin{eqnarray}
{\mathcal M}_{ij}(x,k_\perp )  &=& \frac{1}{2N_c}\left[ f_1(x,k_\perp)\gamma^- - f_{1T}^\perp(x,k_\perp)\gamma^- k_{\perp}\cdot \tilde S_{\perp}\frac{1}{M} + \biggr ( g_{1L}(x,k_\perp) S_L   \right.\nonumber\\
&& -g_{1T}(x,k_\perp) k_\perp\cdot S_{\perp}\frac{1}{M} \biggr ) \gamma_5\gamma^-+h_{1T}(x,k_\perp)i\sigma^{-\mu}\gamma_5S_{\perp \mu}+h_{1}^\perp(x,k_\perp)\sigma^{\mu-}k_{\perp\mu}\frac{1}{M}    \nonumber\\
&&
+ \biggr ( h_{1L}^\perp(x,k_\perp) S_L \frac{1}{M} \left. -h_{1T}^\perp(x,k_\perp)k_\perp\cdot S_\perp \frac{1}{M^2}  \biggr )  i\sigma^{-\mu}\gamma_5 k_{\perp\mu} \right]_{ij}
\label{DECM}
\end{eqnarray}
where $\tilde S^\mu_\perp =\epsilon_\perp^{\mu\nu}S_{\perp\nu}$ with $\epsilon_\perp^{\mu\nu} =\epsilon^{\alpha\beta\mu\nu}l_\alpha n_\beta$.
At leading power there are 8 TMD quark distributions. Among them $f_{1T}^\perp$ and $h_{1}^\perp$ will be zero if T-odd effects are absent 
in the matrix element or the scattering amplitude with the operator in Eq.(\ref{DENM}) has no absorptive parts. In general, such T-odd effects 
of QCD are expected to be nonzero.  

\par 
The definitions of  TMD quark fragmentation functions are conveniently given in a frame where the produced 
hadron moving in the $-z$-direction. The density matrix for fragmentation is defined as :
\begin{eqnarray} 
    \Delta_{ij} (z, k_\perp) = \frac{1}{4z} \int \frac{ d\xi^+ d^2\xi_\perp}{(2\pi)^3} e^{i k\cdot\xi } 
             \sum_X \langle 0\vert q_i (\xi) \vert h(P_h, S_h) X\rangle \langle X h(P_h,S_h) \vert \bar q_j (0) \vert0 \rangle \biggr\vert_{\xi^-=0}, 
\label{DEND} 
\end{eqnarray}               
where the produced hadron moves in the $-z$-direction. The quark which decays into the hadron $h$ carries the $-$- component 
of the momentum as $k^- = P_h^-/z$ and the transverse momentum $k_\perp^\mu$, which is related to the covariantly defined $P_{h\perp}^\mu$
in Eq.(\ref{TD}) as $k_\perp^\mu=-P_{h\perp}^\mu/z$.  
The polarization of the produced hadron is given by the spin vector $S_{h}$ which can be parameterized as\cite{TMDMT}:
\begin{equation} 
 S_h^\mu = S_{hL} \frac{ P_h \cdot l n^\mu -P_h\cdot n l^\mu}{M_h} + S^\mu_{h\perp}, \ \ \ \  S_{h L} = M_h\frac{l\cdot S_h}{l\cdot P_h}
\end{equation} 
with $M_h$ as the hadron mass. 
The density matrix can be decomposed with various TMD fragmentation functions:
\begin{eqnarray}
 \Delta_{ij}(x,k_\perp )  &=& \frac{1}{4}\biggr [ D_1(z, k_\perp)\gamma^+  -  D_{1T}^\perp(z ,k_\perp)\gamma^+ k_{\perp}\cdot \tilde S_{h\perp}\frac{1}{M_h} + \biggr ( G_{1L}(z,k_\perp) S_{hL}  
 \nonumber\\
&& -G_{1T}(z,k_\perp) k_\perp\cdot S_{h\perp}\frac{1}{M_h}\biggr ) \gamma_5\gamma^+ +H_{1T}(z,k_\perp)i\sigma^{+\mu}\gamma_5S_{h\perp \mu}+H_{1}^\perp(z,k_\perp)\sigma^{\mu +}k_{\perp\mu}\frac{1}{M_h}   
 \nonumber\\
&&
+\biggr ( H_{1L}^\perp(z,k_\perp) S_{hL} \frac{1}{M_h}  -H_{1T}^\perp(z,k_\perp) k_\perp\cdot S_{h\perp} \frac{1}{M_h^2} \biggr )  i\sigma^{+ \mu}\gamma_5 k_{\perp\mu} \biggr ]_{ij}. 
\label{DECF}
\end{eqnarray}
There are eight fragmentation functions at leading power. 

In the above definitions of TMD quark distributions and TMD quark fragmentation functions, we have ignored the gauge links. These links are important for making the definitions gauge invariant. We will come back to the issue of gauge invariance at the end of the section. 
The differential cross-section of the SIDIS at tree-level of QCD is represented by the diagram in Fig.2 and the complex conjugated 
diagram. It can be written in the form:
\begin{eqnarray}
d\sigma &=& \frac{1}{2 S_{\ell P} } \frac{d^3 k_2}{2 k_2^0(2\pi)^3} \frac{d^3 P_h}{2 P_h^0(2\pi)^3} \int d^4 p d^4 k \left [  \Gamma^{(L)}_{m i} (p,k) 
   \Gamma_{jn}^{(R)} (p,k) (2\pi)^4 \delta^4 ( p+q -k)  \right ]
 \nonumber\\
     &&  \int \frac{ d^4 x } {(2\pi)^4} e^{-ix\cdot p}
\langle P(P,S) \vert   \bar q_j (x) 
  q_i  (0) \vert P(P,S) \rangle
\nonumber\\  
   && \int \frac{ d^4 y }{(2\pi)^4}  e^{i k\cdot y } 
             \sum_X \langle 0\vert q_n (y ) \vert h(P_h, S_h) X\rangle \langle X h(P_h,S_h) \vert \bar q_m (0) \vert0 \rangle +h.c. ,  
\label{DFCR} 
\end{eqnarray}              
where the quark density matrix in the second line is represented by the grey box in the bottom of Fig.2., and the density matrix for fragmentation in the third line is represented by the grey box in the middle of Fig2.  $\Gamma^{(L)}$ represents the part outside the boxes left to the cut, which is essentially the $\ell +q^*$ scattering amplitude through two photon exchange, while $\Gamma^{(R)}$
is the part right to the cut, which is essentially the $\ell +q^*$ scattering amplitude through one photon exchange. 
$i,j,m$ and $n$ are color- and Dirac indices for the corresponding quark lines given in Fig.2., respectively.  Before the approximation discussed in the below, the initial quark $q^*$ represented by the quark line leaving or entering the grey box in the bottom is off-shell in general. 

%%%%%%%%%%%%%%%% Inset Fig. 1 here %%%%%%%%%%%%%%%%%%%%%%%%%%%%%%
\begin{figure}[hbt]
\begin{center}
\includegraphics[width=9cm]{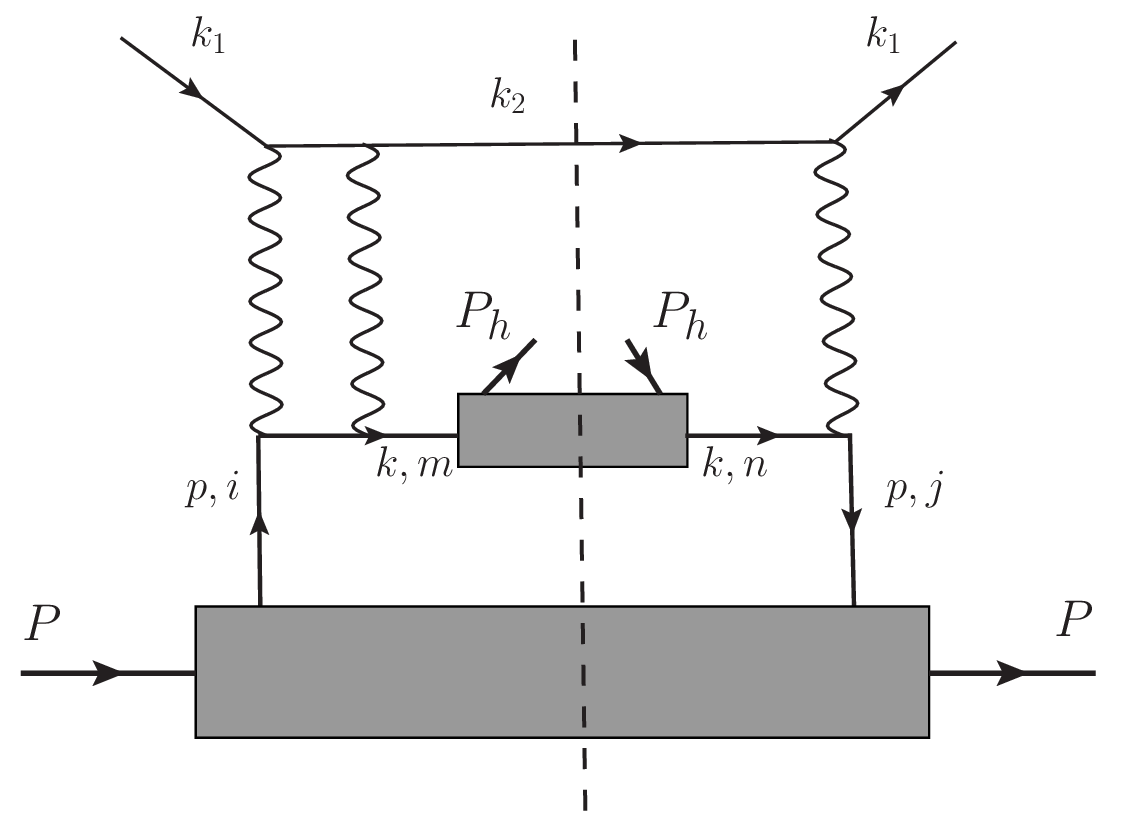}
\end{center}
\caption{Diagrams for the differential cross section discussed in the text. The broken line is the cut.  }
\label{F2}
\end{figure}

%%%%%%%%%%%%%%%% End Fig. 1 here %%%%%%%%%%%%%%%%%%%%%%%%%%%%%%

The frame mentioned before Eq.(\ref{TD}) is not convenient to perform the factorization. It is convenient in the frame in which 
the initial hadron moves in the $+z$-direction, and the final hadron moves in $-z$-direction. In this frame, both hadrons have no transverse momentum transversely to the $z$-direction, but the vector $q^\mu$ does. It has a transverse part given by:
\begin{equation} 
   z_h q_\perp^\mu = -P_{h\perp}^\mu +{\mathcal O}(1/Q), 
\end{equation} 
with $P_{h\perp}$ defined in Eq.(\ref{TD}). In this frame the parton momentum $p$ and $k$ have the pattern: 
\begin{equation} 
    p^\mu \sim Q(1,\lambda^2, \lambda,\lambda), \quad k^\mu \sim Q (\lambda^2, 1, \lambda,\lambda), 
\end{equation} 
with $\lambda = \Lambda_{QCD}/Q$ or  $\lambda = P_{h\perp}/Q$.  Therefore, $\lambda$ is a small parameter for large $Q$.
We can expand the quantities in the $[\cdots]$ of Eq.(\ref{DFCR}) in $\lambda$.  For obtaining the leading order results of the expansion, the first step is to 
neglect $p^-$ and $k^+$ in the $[\cdots]$ in Eq.(\ref{DFCR}). Then the integration over $p^-$ and $k^+$ can be done trivially. After the integration the two density matrices in Eq.(\ref{DFCR}) become ${\mathcal M}$ and $\Delta$, respectively. In the second step one can neglect 
$p_\perp$ and $k_\perp$ in $\Gamma^{(R,L)}$. The transverse parton momenta can not be neglected  in the $\delta$-function, 
because the $\delta$-function also depends on $q_\perp$ or $P_{h\perp}$ which are the same order of $p_\perp$ and $k_\perp$. 
The leading power result for the differential cross section is then 
\begin{eqnarray} 
 d\sigma &=&  \frac{1}{2 S_{\ell P} } \frac{d^3 k_2}{2 k_2^0(2\pi)^3} \frac{d^3 P_h}{2 P_h^0(2\pi)^3}\int d p^+ d^2 p_\perp  d k^- d^2 k_\perp  4z  {\rm Tr}  \biggr [  \Delta (z,k_\perp)  
\Gamma^{(L)} (\hat p, \hat k)  {\mathcal M}(x, p_\perp) 
   \Gamma^{(R)} (\hat p, \hat k) \biggr ] 
\nonumber\\
     && (2\pi)^4 \delta ( p^+ +q ^+) \delta (k^- - q^-) \delta^2 ( p_\perp -k_\perp - P_{h\perp}/z) 
+ h.c.
\nonumber\\
      &=& \frac{1}{2 S_{\ell P} } \frac{d^3 k_2}{2 k_2^0(2\pi)^3} \frac{d^3 P_h}{2 P_h^0(2\pi)^3}\int  d^2 p_\perp  d^2 k_\perp  4z_h  {\rm Tr}  \biggr [  \Delta (z_h, k_\perp)  
\Gamma^{(L)} (\hat p, \hat k)  {\mathcal M}(x_B, p_\perp) 
   \Gamma^{(R)} (\hat p, \hat k) \biggr ] 
\nonumber\\
     && (2\pi)^4   \delta^2 ( p_\perp -k_\perp - P_{h\perp}/z_h) 
+ h.c.
\label{DFCRL} 
\end{eqnarray} 
with $\hat p$ and $\hat k$ as:
\begin{equation}
   \hat p^\mu = (p^+,0,0,0) = (x P^+, 0,0,0), \quad \hat k^\mu =(0, k^-, 0,0)= (0,P_h^-/z, 0,0). 
\end{equation} 
These momenta are those of on-shell partons.  With these on-shell momenta $\Gamma^{(R,L)} (\hat p, \hat k)$ are related to the scattering amplitudes as:
\begin{equation}            
 {\mathcal M}_{2\gamma}(\hat p, \hat k) = \bar u(\hat k) \Gamma^{(L)}(\hat p, \hat k) u(\hat p), \quad  {\mathcal M}^\dagger_{1\gamma}(\hat p, \hat k) =\bar u(\hat p) \Gamma^{(R)} (\hat p, \hat k) u(\hat k). 
 \label{GALR}
\end{equation}   
The detailed results for $\Gamma^{(R,L)} (\hat p, \hat k)$ can be read from results in the last section. The transverse 
metric and anti-symmetic tensor in this frame are 
\begin{equation}            
g_\perp^{\mu\nu}=g^{\mu\nu}-\frac{\hat{p}^\mu\hat{k}^\nu+\hat{p}^\nu \hat{k}^\mu}{\hat{p}\cdot\hat{k}},\quad
\epsilon_\perp^{\mu\nu}=\epsilon^{\mu\nu\rho\sigma}\hat{p}_\rho\hat{k}_\sigma\frac{1}{\hat{p}\cdot\hat{k}}.
\end{equation}

Taking the results for the absorptive part of the two-photon exchange amplitude, we can evaluate the trace in Eq.(\ref{DFCRL}). We find 
that there is no contribution to the differential cross section when the initial lepton is unpolarized. When the lepton is polarized with the helicity $\lambda_\ell$,  the differential cross-section only receives contributions from  the chirality-odd 
part of TMD parton distributions and TMD quark fragmentation functions. We have the sum of the trace and its complex conjugated part:
\begin{align}
&4z_h \text{Tr}\Big[\Delta(z_h,k_\perp)\Gamma^{(L)}(\hat{p},\hat{k})
\mathcal{M}(x_B,p_\perp)\Gamma^{(R)}(\hat{p},\hat{k})
\Big]   +h.c. \notag\\
=& \frac{2 e^6e_q^3 z_h}{t^2}
\Big\{
 h_{1T}(x_B ,p_{\perp}^2)\Big[-H_{1T}(z_h,k_{\perp}^2) F_3(S_\perp,S_{h\perp})
-H_{1s}^\perp(z_h,k_{\perp}) F_3(S_\perp,\frac{k_{\perp}}{M_h})
\notag\\
&+H_1^\perp(z_h,k_{\perp}^2)F_4(S_\perp,\frac{k_{\perp}}{M_h})
\Big]+h_{1s}^\perp(x_B,p_{\perp})\Big[-H_{1T}(z_h,k_{\perp}^2) F_3(\frac{p_{\perp}}{M},S_{h\perp})
-H_{1s}^\perp(z_h,k_{\perp}) F_3(\frac{p_{\perp}}{M},\frac{k_{\perp}}{M_h})
\notag\\
& +H_{1}^\perp(z_h,k_{\perp}^2) F_4(\frac{p_{\perp}}{M},\frac{k_{\perp}}{M_h})\Big]
+h_1^\perp(x_B,p_{\perp}^2)\Big[-
H_{1T}(z_h,k_{\perp}^2) F_4(\frac{p_{\perp}}{M},S_{h\perp})
-H_{1s}^\perp(z_h,k_{\perp}) F_4(\frac{p_{\perp}}{M},\frac{k_{\perp}}{M_h})
\notag\\
& -H_{1}^\perp (z_h,k_{\perp}^2)F_3(\frac{p_{\perp}}{M},\frac{k_{\perp}}{M_h})
\Big]
\Big\},
\end{align} 
with the combinations $H_{1s}^\perp$ and $h_{1s}^\perp$ as
\begin{eqnarray}
H_{1s}^\perp(z,k_\perp) &=& H_{1L}^\perp(z,k_\perp) S_{hL}  -H_{1T}^\perp(z,k_\perp) k_\perp\cdot S_{h\perp} \frac{1}{M_h},
\nonumber\\
 h_{1s}^\perp(x, p_\perp) &=& h_{1L}^\perp(x,k_\perp) S_L  -h_{1T}^\perp(x,k_\perp)k_\perp\cdot S_\perp \frac{1}{M} .  
\end{eqnarray} 
The two functions $F_{3,4}$ with two transverse vectors $h_1$ and $h_2$ as variables are  defined as
\begin{align}
F_3(h_1,h_2)=&  4\lambda_\ell \hat p \cdot \hat k \Big[
{\text Im} ({\mathcal A}_2 t+4{\mathcal A}_3 s+24 {\mathcal A}_4)(h_2\cdot k_1 \epsilon^{h_1k_1}_\perp+h_1\cdot k_1 \epsilon_\perp^{h_2k_1})\notag\\
&+2{\text Im }{\mathcal A}_3 t (h_2\cdot k_1 \epsilon_\perp^{h_1 k_1}-h_1\cdot k_1 \epsilon_\perp^{h_2 k_1})
-2{\text Im} {\mathcal A}_3 s(s+t)\epsilon_\perp^{h_1 h_2}\Big],\notag\\
\ 
F_4(h_1,h_2)=&\lambda_\ell \Big[{\text Im}({\mathcal A}_2 t+4{\mathcal A}_3 s+24 {\mathcal A}_4 )\Big]
(-2s(s+t)h_1\cdot h_2 +4t h_1\cdot k_1 h_2\cdot k_1),
\end{align}
where ${\text Im} {\mathcal A}_{2,3,4}$ are given in Eq.(\ref{ImA}). We notice that the divergent ${\text Im}{\mathcal A}_1$ does not give contribution here. Therefore, the contribution of  T-odd effect considered here is finite.  The transverse vectors $h_1$ and $h_2$ are transverse to $\hat p$ and $\hat k$.  We use the short notation $\epsilon_\perp^{h_1h_2}$ for $\epsilon_\perp^{\mu\nu} h_{1\mu} h_{2\nu}$.  The result in the above are invariant. 

The results of the two functions $F_{3,4}$ can be simplified if we introduce the transverse vector $k_{1\perp}^\mu =g_\perp^{\mu\nu} k_{1\nu}$
and use the results of ${\text Im} {\mathcal A}_{2,3,4}$. The result is:
\begin{eqnarray}
   F_3(h_1,h_2) &=& \lambda_\ell  \frac{1}{2\pi} \frac{ s t }{(s+t)^2} C_1 (y) h_2\cdot k_{1\perp} ( \epsilon_\perp^{h_1  k_{1\perp} }+ \epsilon_\perp^{h_2 k_{1\perp}} ),  
\nonumber\\
    F_4 (h_1,h_2) &=&\lambda_{\ell} \frac{1}{2\pi} \frac{s^2}{s+t} C_1 (y)( h_1\cdot h_2 -2 \frac{t }{s (s+t)}  h_1\cdot  k_{1\perp} h_2\cdot k_{1\perp} ),  
\end{eqnarray}  
 with $C_1(y)= 1-y+(2-y) \ln y$. 
\par
Our results are derived from the diagram in Fig.2. At leading order there are contributions from gluons exchanged between 
quark lines and grey boxes and between grey boxes. At leading power these gluons are collinear gluons and soft gluons. 
The contributions from collinear gluons can be summed with gauge links in the standard definitions of TMD quark distributions and fragmentation functions. This makes them gauge invariant. The contributions for soft gluons are summed into a soft factor. From 
Eq.(\ref{GALR}) $\Gamma^{(L,R)}$ are extracted from on-shell scattering amplitudes and they are gauge invariant. Therefore, our results for the differential cross section derived here are gauge invariant. 

It is noted that the contributions from two-photon exchange are not only from the diagram given in Fig.2, they can be from other diagrams, where one photon in the left part of Fig.2 is absorbed to one of the two grey boxes instead connecting to the struck quark.  In the case that the photon is absorbed by the lower grey box, it is expected that the contribution can partly be represented by inserting QED gauge links into the definition 
of TMD quark distributions in Eq.(\ref{DENM}). This makes the definition QED gauge invariant. The remaining 
contribution can be represented by matrix elements like $\langle P\vert \bar q F^{+\mu} q\vert P\rangle$ and are power suppressed.  At the leading power, the contribution in this case does not give any new effect, since the perturbative coefficient functions are still determined by one-photon exchange. This happens also to the case if one photon is absorbed by the grey box representing the fragmentation.

\par\vskip20pt
\noindent 
{\bf 4. Angular Distributions}

From our results in the last section, one can derive angular distributions in various frames. In this section 
we derive the angular distribution of T-odd effect in a frame specified in the following:  In the frame, the virtual photon 
moves in $-z$-direction and the initial hadron moves in the $+z$-direction. 
The initial- and final lepton spans the so-called lepton plane. The azimuthal angle between the transverse spin vector $S_\perp$ and the lepton plane is denoted by $\phi_s$. Similarly, one defines the azimuthal angle $\phi_h$ for the produced hadron.
The azimuthal angle of the outgoing lepton around the lepton beam with respect to the spin vector is denoted by $\psi_\ell$.
In the kinematical region of SIDIS, one has $d\psi_\ell\approx d\phi_s$\cite{Diehl}.

For simplicity, we will consider here only the case that the hadron in the final state is unpolarized. The differential cross section for the T-odd effect in the given frame is given by:
\begin{align}
\frac{d\sigma}{d\phi_s dQ^2 dy dz_h d^2 P_{h\perp}}
=& \lambda_l \frac{\alpha^3 e_q^3 x_B}{ Q^4}\frac{C_1(y)}{1-y}
\Big\{
-\mathcal{C}\Big[w_3 h_{1}^\perp H_1^\perp\Big]\sin (2 {\phi}_h)
-S_L \mathcal{C}\Big[w_3 h^\perp_{1L} H_1^\perp\Big]\cos(2 {\phi}_h) \notag\\
&+|\vec{S}_\perp|\Big[-\Big(\mathcal{C}\Big[w_1 h_{1T} H_1^\perp\Big]
+\frac{1}{2}\mathcal{C}\Big[w_6 h^\perp_{1T} H_1^\perp\Big]
\Big) \cos(3{\phi}_h-{\phi}_s) \notag\\ 
     &+\frac{1}{2}\mathcal{C}\Big[w_4 h^\perp_{1T}H_1^\perp\Big]\cos({\phi}_h+{\phi}_s)\Big]\Big\}.
\end{align}
The convolution in transverse momentum space is defined by 
\begin{align}
\mathcal{C}\Big[w F(x,p_\perp^2)D(z,k_\perp^2)\Big]
=\int d^2 p_\perp d^2 k_\perp \delta^2(p_\perp+q_\perp-k_\perp)
w(p_\perp,k_\perp) F(x,p_\perp^2)D(z,k_\perp^2).
\label{DFCS}
\end{align}
The weights appearing in differential cross sections are 
\begin{align}
\notag
w_1=& \frac{k_\perp\cdot\hat{q}_\perp}{M_h},\quad 
w_2= \frac{p_\perp\cdot k_\perp}{M M_h},\quad
w_3=\frac{2p_\perp\cdot \hat{q}_\perp k_\perp\cdot \hat{q}_\perp+p_\perp\cdot k_\perp}{M M_h},
\quad 
w_6=  -\frac{p_\perp^2 k_\perp\cdot \hat{q}_\perp}{M^2 M_h}
\\ 
w_4=& -\frac{k_\perp\cdot\hat{q}_\perp [4(p_\perp\cdot\hat{q}_\perp)^2+p_\perp^2]
+2p_\perp\cdot k_\perp p_\perp\cdot\hat{q}_\perp}{M^2 M_h},\quad
w_5= -\frac{k_\perp\cdot\hat{q}_\perp(p_\perp\cdot\hat{q}_\perp)^2+p_\perp\cdot\hat{q}_\perp p_\perp\cdot k_\perp}{
M^2 M_h}, 
\end{align}
with $\hat q_\perp^\mu = -P_{h\perp}^\mu/\vert P_{h\perp}\vert$. 

It is interesting to compare the differential cross section of the T-odd effect in Eq.(\ref{DFCS}) with that obtained from one-photon exchange approximation. One finds that the part of the angular distribution proportional to the lepton helicity has the following angular dependence:
\begin{equation}
\frac{d\sigma}{d\phi_s dQ^2 dy dz_h d^2 p_{h\perp}} \propto
   \lambda_\ell \biggr [  a_1 \sin\phi_h +S_L (a_2 +a_3 \cos\phi_h ) + \vert S_\perp\vert (a_4 \cos(\phi_h -\phi_s) + a_5 \cos\phi_s+a_6 \cos(2\phi_h+\phi_s)) \biggr ],  
 \end{equation} 
 where the coefficients $a_i(i=1,2,\cdots 6)$ can be extracted from the results in \cite{BDGM}. Comparing this with Eq.(\ref{DFCS}), it is clearly 
 that there are additional angular dependences introduced by the T-odd effects. 

Based on our results one can define various asymmetries to measure the studied T-odd effects.  An asymmetry can be defined in general as:
\begin{align}
A_L[f ]=& \frac{\int_0^{2\pi} d\phi_h  f 
[d\sigma(\lambda_\ell)-d\sigma (-\lambda_\ell)]}{\int_0^{2\pi} d\phi_h [d\sigma (\lambda_\ell)
+d\sigma (-\lambda_\ell)]}.
\end{align}
with the function $f$ depending on $\phi_h$ as the weight. We can define 6 asymmetries in the following. These asymmetries are zero if the T-odd effects are neglected. With our results we have: 
\begin{align}
\notag
A_L[\cos 2{\phi}_h]=& \lambda_\ell \frac{\alpha e_q^3 }{2}\frac{y C_1(y)}{(1-y)(y^2-2y+2)}\frac{1}{e_q^2 \mathcal{C}[f_1D_1]}
S_L\Big( \mathcal{C}[w_3 h_{1L}^\perp H_1^\perp]\Big),\notag\\
A_L[\sin 2{\phi}_h]=& \lambda_\ell \frac{\alpha e_q^3 }{2}\frac{y C_1(y)}{(1-y)(y^2-2y+2)}\frac{1}{e_q^2 \mathcal{C}[f_1D_1]}
\Big(-\mathcal{C}[w_3 h_{1}^\perp H_1^\perp]\Big),\notag\\
A_L[\cos 3{\phi}_h]=& \lambda_\ell \frac{\alpha e_q^3 }{2}\frac{y C_1(y)}{(1-y)(y^2-2y+2)}\frac{1}{e_q^2 \mathcal{C}[f_1D_1]}
|\vec{S}_{\perp}|\cos{\phi}_s
\Big( \mathcal{C}[w_1 h_{1T} H_1^\perp]+\frac{1}{2}\mathcal{C}[w_6 h_{1T}^\perp H_1^\perp]\Big),\notag\\
A_L[\sin 3{\phi}_h]=& -\lambda_\ell \frac{\alpha e_q^3 }{2}\frac{y C_1(y)}{(1-y)(y^2-2y+2)}\frac{1}{e_q^2 \mathcal{C}[f_1D_1]}
|\vec{S}_{\perp}|\sin{\phi}_s
\Big( \mathcal{C}[w_1 h_{1T} H_1^\perp]+\frac{1}{2}\mathcal{C}[w_6 h_{1T}^\perp H_1^\perp]\Big),\notag\\
A_L[\cos{\phi}_h]=& \lambda_\ell \frac{\alpha e_q^3 }{2}\frac{y C_1(y)}{(1-y)(y^2-2y+2)}\frac{1}{e_q^2 \mathcal{C}[f_1D_1]}
|\vec{S}_{\perp}|\cos{\phi}_s
\Big( -\frac{1}{2}\mathcal{C}[w_4 h_{1T}^\perp H_1^\perp]\Big),\notag\\ 
A_L[\sin{\phi}_h]=& -\lambda_\ell \frac{\alpha e_q^3 }{2}\frac{y C_1(y)}{(1-y)(y^2-2y+2)}\frac{1}{e_q^2 \mathcal{C}[f_1D_1]}
|\vec{S}_{\perp}|\sin{\phi}_s
\Big( \frac{1}{2}\mathcal{C}[w_4 h_{1T}^\perp H_1^\perp]\Big).
\end{align} 
These asymmetries are zero if one neglects the absorptive part from two-photon exchange.  The results in the above are given  for  the case 
there is one quark flavor. A  sum of flavors should be performed for the case of many flavors. 
A result about one slightly different asymmetry 
about the T-odd effect is given 
in \cite{Schlegel:2009pw}. It corresponds to the second asymmetry in the above. With our results an agreement can be found.

\par\vskip20pt

\noindent 
{\bf 5. Summary} 

We have studied the T-odd effects from QED one-loop correction in SIDIS at low transverse momentum. The T-odd effects 
come from the scattering of the initial lepton with a quark from the initial proton where two photons are exchanged between the lepton 
and the quark. The absorptive part of the scattering amplitude is calculated. The result has an I.R. divergence.  After performing TMD factorization 
for the process, the differential cross section is obtained  as convolutions of TMD quark distributions and TMD quark fragmentation 
functions, and it is finite.   The studied T-odd effects only appear when the initial lepton is polarized. It adds extra angular dependence in the differential cross section. We have derived the angular distribution in the case where the hadron in the final state is unpolarized. Six asymmetries are constructed,  
they are determined by the absorptive part.  Our results presented here are important not only for detecting physical effects in the process beyond one-photon exchange, but also for searching for new physics.  

\par\vskip60pt
\begin{center}
{\Large\bf Appendix}  
\end{center} 
\renewcommand{\theequation}{A.\arabic{equation}}
\setcounter{equation}{0}
In this Appendix we present our results for the amplitude of the two-photon exchange in detail. 
From the amplitude of two photon exchange in Eq.(\ref{2PE}), we can define the following loop integrals in $d$-dimension:
\begin{align}
\{I_0,K^\mu,J^{\mu\nu}\}\equiv -i\int\frac{d^d l}{(2\pi)^d}
\frac{\{1,l^\mu,l^\mu l^\nu\}}{l^2(l+k_1-k_2)^2(l+k_1)^2(l-p_1)^2}.
\end{align}
For simplicity, $+i\epsilon$ in the denominator of propagators is suppressed. 

In physical region where $s=(k_1+p_1)^2>0,t=(k_1-k_2)^2<0$, these integrals contain nonzero imaginary parts. 
For scalar integral, the result is 
\begin{align}
\notag
I_0(s,t)=& \frac{K_\epsilon}{st}\Big[
\frac{16}{\epsilon^2}-\frac{4 (\ln (s)+\ln (-t)+2)}{\epsilon }+\left(2 \ln (s) (\ln
   (-t)+1)+2 \ln (-t)-\frac{5 \pi ^2}{3}+4\right)\\ 
& +i\pi\Big( \frac{4}{\epsilon } -2 \ln (-t)-2  \Big) +O\left(\epsilon ^1\right)
\Big].
\end{align}  
This result agrees with the result in e.g.,\cite{Ellis:2007qk}. 

The vector integral $K^\alpha$ are decomposed into three scalar integrals $K_{1,2,3}$, 
\begin{align}
K^\alpha=& k_1^\alpha K_1(s,t) + k_2^\alpha K_2(s,t)+p_1^\alpha K_3(s,t).
\end{align}
The results are 
\begin{align}
\notag
K_1(s,t)=& -\frac{1}{2}I_0(s,t),\notag\\ 
K_2(s,t)=& \frac{K_\epsilon}{st}\Big[
\frac{4}{\epsilon ^2}-\frac{2 (\ln (s)+1)}{\epsilon }+i\pi \Big(\frac{2}{\epsilon }-\frac{s \ln (-t)+t \ln (s)+s+t}{s+t}\Big)     \notag\\
   &+\frac{3 t \ln
   ^2(s)-3 s \ln ^2(-t)+6 \ln (s) (s \ln (-t)+s+t)-4 \pi ^2 s+6 s-\pi ^2
   t+6 t}{6 (s+t)}
\Big],\notag\\
K_3(s,t)=&\frac{K_\epsilon}{st}\Big[
\frac{4}{\epsilon ^2}-\frac{2 (\ln (-t)+1)}{\epsilon }+i\pi\frac{t\ln\frac{s}{-t}}{s+t} \notag\\ 
&+\frac{-3 t \ln
   ^2(s)+3 s \ln ^2(-t)+6 \ln (-t) (t \ln (s)+s+t)-\pi ^2 s+6 s-4 \pi ^2
   t+6 t}{6 (s+t)}   \Big].
\end{align}
The tensor integral $J^{\alpha\beta}$ is decomposed into seven scalar integrals:
\begin{align}
J^{\alpha\beta}=& J_0(s,t) g^{\alpha\beta}+J_1(s,t) k_1^\alpha k_1^\beta + J_2(s,t) (k_1^\alpha k_2^\beta+k_1^\beta k_2^\alpha)
+J_3(s,t)(k_1^\alpha p_1^\beta+k_1^\beta p_1^\alpha)
+J_4(s,t) k_2^\alpha k_2^\beta    \notag\\
&+J_5(s,t) (k_2^\alpha p_1^\beta+k_2^\beta p_1^\alpha)
+J_6(s,t) p_1^\alpha p_1^\beta.
\end{align}
The results of the scalar functions are 
\begin{align}
J_0(s,t)=& K_\epsilon\Big[\frac{(\ln (s)-\ln (-t))^2}{4 (s+t)}+i\pi \frac{\ln (-t)-\ln (s)}{2 (s+t)} \Big],\notag\\
J_1(s,t)=& \frac{K_\epsilon}{st}\Big[
\frac{8}{\epsilon ^2}-\frac{2 (\ln (s)+\ln (-t))}{\epsilon }+\left(\ln (s) \ln
   (-t)-\frac{5 \pi ^2}{6}+4\right)  +i\pi\Big( \frac{2}{\epsilon }-\ln (-t)  \Big)
\Big],\notag\\
J_2(s,t)=& \frac{K_\epsilon}{st}\Big[
-\frac{4}{\epsilon ^2}+\frac{2 \ln (s)}{\epsilon }
+\frac{4 \left(\pi^2-3\right) (s+t)-3 t \ln ^2(s)+3 s \ln ^2(-t)-6 s \ln (s) 
\ln(-t)}{6 (s+t)}\notag\\
   &+i\pi\Big( -\frac{2}{\epsilon }+\frac{t \ln (s)+s \ln (-t)}{s+t}\Big) \Big],\notag\\
J_3(s,t)=& \frac{K_\epsilon}{st}\Big[\frac{2}{\epsilon }-\ln (-t)+1\Big],\notag\\
J_4(s,t)=& \frac{K_\epsilon}{st}\Big[
\frac{4}{\epsilon ^2}+\frac{2-2 \ln (s)}{\epsilon }   + \frac{1}{6(s+t)^2} \Big ( -3 s^2 \ln^2(-t)+6 \ln (s) \left(s^2 \ln (-t)-t (s+t)\right) 
 \notag\\ 
 & -2 \left(2 \pi^2-9\right) (s+t)^2+3 t (2 s+t) \ln ^2(s)-6 s (s+t) \ln (-t) \Big ) 
 \notag\\
   &+i\pi \Big( 
\frac{2}{\epsilon }+\frac{s^2 (-\ln (-t))+t (s+t)-t (2 s+t) \ln
   (s)}{(s+t)^2}
\Big) \Big],\notag\\
J_5(s,t)=& \frac{K_\epsilon}{st}\Big[
-\frac{2}{\epsilon }-\frac{2 (s+t)^2+s t \ln ^2(s)+s t \ln ^2(-t)-2 s
   (s+t) \ln (-t)-2 t \ln (s) (s \ln (-t)+s+t)}{2
   (s+t)^2}\notag\\
&+i\pi \Big(
-\frac{t (s \ln (-t)+s+s (-\ln (s))+t)}{(s+t)^2}
\Big) \Big] ,\notag\\
J_6(s,t)=& \frac{K_\epsilon}{st}\Big[
\frac{4}{\epsilon ^2}+\frac{2-2 \ln (-t)}{\epsilon }- \frac{1}{6 (s+t)^2} \Big ( 
 3 t^2 \ln^2(s)
+\left(\pi ^2-18\right) (s+t)^2-3 s (s+2 t) \ln ^2(-t)  \notag\\
   & +6 s (s+t)
   \ln (-t)+6 t \ln (s) (s+t-t \ln (-t)) \Big ) 
  +i\pi \Big(\frac{t (t \ln (s)+s+t-t \ln (-t))}{(s+t)^2}
\Big) \Big].
\end{align}

$\mathcal{A}_i$ are defined in Eq.(\ref{ADEF}). These coefficient functions are can be obtained from the results 
of loop integrals. They are: 
\begin{align}
\mathcal{A}_1=& 2s\Big[I_0-2J_2-J_3+J_5+K_1-K_2-K_3\Big],\notag\\
\mathcal{A}_2=&8\Big[J_2-J_5+K_2\Big],\notag\\
\mathcal{A}_3=&-J_3-J_5,\notag\\
\mathcal{A}_4=& -J_0.
\end{align} 
With the known results of scalar integrals, we have 
\begin{align}
\mathcal{A}_1=& K_\epsilon\Big(\frac{\mu^2}{-t}\Big)^{\epsilon/2}\frac{1}{t} 
\Big\{\frac{16}{\epsilon ^2}+\frac{8 (\ln (y)-1)}{\epsilon }+\frac{-4 \left(2
   \pi ^2-3\right) (y-1)^2+3 y (2 y-1) \ln ^2(y)-6 y (y-1) \ln (y)}{3
   (y-1)^2} \notag\\
&+i\pi\Big[\frac{8}{\epsilon }+\frac{2 y (-y+(2 y-1) \ln
   (y)+1)}{(y-1)^2} \Big]\Big\},\notag\\
\mathcal{A}_2=&K_\epsilon\Big(\frac{\mu^2}{-t}\Big)^{\epsilon/2}\frac{1}{t^2} 
\Big\{
\frac{4y(2-2y+y\ln y)}{(y-1)^2}[\ln y +i\pi]
\Big\},\notag\\
\mathcal{A}_3=&K_\epsilon\Big(\frac{\mu^2}{-t}\Big)^{\epsilon/2}\frac{y^2}{t^2(y-1)^2} 
\Big\{
\ln y(1-y+\frac{1}{2}\ln y)+i\pi (1-y+\ln y)
\Big\},\notag\\
\mathcal{A}_4=&K_\epsilon\Big(\frac{\mu^2}{-t}\Big)^{\epsilon/2}\frac{y^2}{t^2(y-1)^2} 
\Big\{
\frac{y\ln y (\frac{1}{2}\ln y +i\pi)}{2t(1-y)}
\Big\},
\end{align}
where we have introduced the variable $y=-t/s$.  At leading power of TMD factorization we have  $p_1=x_B P$. Then  $y$ 
is equal to the one defined at hadron level.

\par\vskip40pt
% The correct funding number:  12075299 (mian shang 2021-2024) 
% qunti 11821505, penghuanwu 11847612
% Adding lattice funding number 11935017(2020-2024)
% The number 11675241 is for 1017-2020. 

\noindent
{\bf Acknowledgments}
\par
The work is supported by  National Key R\&D Program of China No. 2024YFE0109800,801.

\par


\vskip40pt


\begin{thebibliography}{99}

% QED corrections

%\cite{Akushevich:2026mje}
\bibitem{Akushevich:2026mje}
I.~Akushevich, H.~Gao, A.~Ilyichev, S.~Jia, V.~Khachatryan, Y.~Lin, T.~Liu and W.~Melnitchouk,
%``QED radiative effects in semi-inclusive deep-inelastic scattering: traditional and factorized approaches,''
[arXiv:2608.20294 [hep-ph]].
%0 citations counted in INSPIRE as of 02 Sep 2026


%\cite{Akushevich:2023jjc}
\bibitem{Akushevich:2023jjc}
I.~Akushevich, H.~Avakian, A.~Ilyichev and S.~Srednyak,
%``Complete lowest order radiative corrections in semi-inclusive scattering of polarized particles,''
Eur. Phys. J. A \textbf{59} (2023), 246
doi:10.1140/epja/s10050-023-01150-0
[arXiv:2310.17961 [hep-ph]].
%3 citations counted in INSPIRE as of 02 Sep 2026

%\cite{Liu:2020rvc}
\bibitem{Liu:2020rvc}
T.~Liu, W.~Melnitchouk, J.~W.~Qiu and N.~Sato,
%``Factorized approach to radiative corrections for inelastic lepton-hadron collisions,''
Phys. Rev. D \textbf{104} (2021) no.9, 094033
doi:10.1103/PhysRevD.104.094033
[arXiv:2008.02895 [hep-ph]].
%35 citations counted in INSPIRE as of 01 Sep 2026%\cite{Cammarota:2025jyr}


\bibitem{Cammarota:2025jyr}
J.~Cammarota, J.~W.~Qiu, K.~Watanabe and J.~Y.~Zhang,
%``Factorized QED and QCD contribution to deeply inelastic scattering,''
Phys. Rev. D \textbf{112} (2025) no.5, 056007
doi:10.1103/2d8y-ljwx
[arXiv:2505.23487 [hep-ph]].
%10 citations counted in INSPIRE as of 01 Sep 2026




\bibitem{CSS} J.C. Collins, D.E. Soper and G. Sterman, Nucl. Phys. B250 (1985) 199, Nucl. Phys. B261, 104 (1985).

\bibitem{Ji:2004wu}
X.~d.~Ji, J.-P.~Ma and F.~Yuan,
%``QCD factorization for semi-inclusive deep-inelastic scattering at low transverse momentum,''
Phys. Rev. D \textbf{71} (2005), 034005
doi:10.1103/PhysRevD.71.034005
[arXiv:hep-ph/0404183 [hep-ph]].
%804 citations counted in INSPIRE as of 12 Aug 2026

%\cite{Boussarie:2023izj}
\bibitem{Boussarie:2023izj}
R.~Boussarie, M.~Burkardt, M.~Constantinou, W.~Detmold, M.~Ebert, M.~Engelhardt, S.~Fleming, L.~Gamberg, X.~Ji and Z.~B.~Kang, \textit{et al.}
%``TMD Handbook,''
[arXiv:2304.03302 [hep-ph]].
%266 citations counted in INSPIRE as of 01 Sep 2026

% New Physics with SIDIS...... 


%\cite{Wen:2023xxc}

%\cite{Huang:2025ljp}
\bibitem{Huang:2025ljp}
Y.~Huang, X.~B.~Tong and H.~L.~Wang,
%``Nucleon Energy Correlators as a Probe of Light-Quark Dipole Operators at the Electron-Ion Collider,''
Phys. Rev. Lett. \textbf{136} (2026) no.13, 131902
doi:10.1103/7y5n-cscx
[arXiv:2508.08516 [hep-ph]].
%14 citations counted in INSPIRE as of 01 Sep 2026



%\cite{Wen:2024nff}
\bibitem{Wen:2024nff}
X.~K.~Wen, B.~Yan, Z.~Yu and C.~P.~Yuan,
%``Transverse spin effects and light-quark dipole moments at lepton colliders,''
Phys. Rev. D \textbf{112} (2025) no.5, 053004
doi:10.1103/32nb-d466
[arXiv:2411.13845 [hep-ph]].
%16 citations counted in INSPIRE as of 01 Sep 2026

%\cite{Wen:2023xxc}
\bibitem{Wen:2023xxc}
X.~K.~Wen, B.~Yan, Z.~Yu and C.~P.~Yuan,
%``Single Transverse Spin Asymmetry as a New Probe of Standard-Model-Effective-Field-Theory Dipole Operators,''
Phys. Rev. Lett. \textbf{131} (2023) no.24, 241801
doi:10.1103/PhysRevLett.131.241801
[arXiv:2307.05236 [hep-ph]].
%25 citations counted in INSPIRE as of 01 Sep 2026


% EiC

%\cite{Accardi:2012qut}
\bibitem{Accardi:2012qut}
A.~Accardi, J.~L.~Albacete, M.~Anselmino, N.~Armesto, E.~C.~Aschenauer, A.~Bacchetta, D.~Boer, W.~K.~Brooks, T.~Burton and N.~B.~Chang, \textit{et al.}
%``Electron Ion Collider: The Next QCD Frontier: Understanding the glue that binds us all,''
Eur. Phys. J. A \textbf{52} (2016) no.9, 268
doi:10.1140/epja/i2016-16268-9
[arXiv:1212.1701 [nucl-ex]].
%2165 citations counted in INSPIRE as of 13 Sep 2026


% Eicc 
%\cite{Anderle:2021wcy}
\bibitem{Anderle:2021wcy}
D.~P.~Anderle, V.~Bertone, X.~Cao, L.~Chang, N.~Chang, G.~Chen, X.~Chen, Z.~Chen, Z.~Cui and L.~Dai, \textit{et al.}
%``Electron-ion collider in China,''
Front. Phys. (Beijing) \textbf{16} (2021) no.6, 64701
doi:10.1007/s11467-021-1062-0
[arXiv:2102.09222 [nucl-ex]].
%615 citations counted in INSPIRE as of 13 Sep 2026

%\cite{Xiao:2026tbs}
\bibitem{Xiao:2026tbs}
B.~W.~Xiao, Y.~Zhao and J.~Zhou,
%``Physics of the Electron{\textendash}Ion Collider in China,''
Prog. Part. Nucl. Phys. \textbf{151} (2026), 104264
doi:10.1016/j.ppnp.2026.104264
[arXiv:2608.11712 [hep-ph]].
%3 citations counted in INSPIRE as of 13 Sep 2026



%\cite{Metz:2006pe}
\bibitem{Metz:2006pe}
A.~Metz, M.~Schlegel and K.~Goeke,
%``Transverse single spin asymmetries in inclusive deep-inelastic scattering,''
Phys. Lett. B \textbf{643} (2006), 319-324
doi:10.1016/j.physletb.2006.11.009
[arXiv:hep-ph/0610112 [hep-ph]].
%38 citations counted in INSPIRE as of 21 Aug 2026


%\cite{Metz:2012ui}
\bibitem{Metz:2012ui}
A.~Metz, D.~Pitonyak, A.~Schafer, M.~Schlegel, W.~Vogelsang and J.~Zhou,
%``Single-spin asymmetries in inclusive deep inelastic scattering and multiparton correlations in the nucleon,''
Phys. Rev. D \textbf{86} (2012), 094039
doi:10.1103/PhysRevD.86.094039
[arXiv:1209.3138 [hep-ph]].
%76 citations counted in INSPIRE as of 02 Sep 2026




%\cite{Schlegel:2012ve}
\bibitem{Schlegel:2012ve}
M.~Schlegel,
%``Partonic description of the transverse target single-spin asymmetry in inclusive deep-inelastic scattering,''
Phys. Rev. D \textbf{87} (2013) no.3, 034006
doi:10.1103/PhysRevD.87.034006
[arXiv:1211.3579 [hep-ph]].
%35 citations counted in INSPIRE as of 02 Sep 2026




%\cite{Schlegel:2009pw}
\bibitem{Schlegel:2009pw}
M.~Schlegel and A.~Metz,
%``Two-Photon Exchange in (Semi-)Inclusive DIS,''
AIP Conf. Proc. \textbf{1149} (2009) no.1, 543-546
doi:10.1063/1.3215707
[arXiv:0902.0781 [hep-ph]].
%7 citations counted in INSPIRE as of 02 Sep 2026



% T-odd effects 
%\cite{Kane:1978nd}
\bibitem{Kane:1978nd}
G.~L.~Kane, J.~Pumplin and W.~Repko,
%``Transverse Quark Polarization in Large p(T) Reactions, e+ e- Jets, and Leptoproduction: A Test of QCD,''
Phys. Rev. Lett. \textbf{41} (1978), 1689
doi:10.1103/PhysRevLett.41.1689
%439 citations counted in INSPIRE as of 01 Sep 2026%\cite{Bernreuther:1992ef}

\bibitem{Bernreuther:1992ef}
W.~Bernreuther, J.~P.~Ma and T.~Schroder,
%``Top quark polarization and $T$ odd spin correlations as tools for testing (non) Standard Model predictions,''
Phys. Lett. B \textbf{297} (1992), 318-326
doi:10.1016/0370-2693(92)91269-F
%47 citations counted in INSPIRE as of 28 Jul 2026





\bibitem{TMDBM} D. Boer and P.J. Mulders, Phys. Rev. D57 (1998) 5780, e-Print: hep-ph/9711485.

\bibitem{TMDGMS} K. Goeke, A. Metz and M. Schlegel, Phys. Lett. B618 (2005) 90, e-print:he-ph/0504130.

\bibitem{BDGM} A. Bacchetta {\it et al.}, JHEP 0702 (2007) 093,
e-Print: hep-ph/0611265.

\bibitem{TMDMT} P.J. Mulders and R.D. Tangerman, Nucl. Phys. B461 (1996) 197, e-Print: hep-ph/9510301,
Nucl. Phys. B484 (1997) 538(E).

\bibitem{Diehl} M. Diehl and S. Sapeta,  Eur.Phys.J. C41 (2005) 515 2005,  
e-Print: hep-ph/0503023. 

%\cite{Ellis:2007qk}
\bibitem{Ellis:2007qk}
R.~K.~Ellis and G.~Zanderighi,
%``Scalar one-loop integrals for QCD,''
JHEP \textbf{02} (2008), 002
doi:10.1088/1126-6708/2008/02/002
[arXiv:0712.1851 [hep-ph]].
%558 citations counted in INSPIRE as of 11 Sep 2026

\end{thebibliography}
\end{document}